\documentclass[
 amsmath,
 amssymb,
 aps, 
 prl,
 reprint
]{revtex4-2}

\usepackage{graphicx}
\usepackage{dcolumn}
\usepackage{bm}
\usepackage{hyperref}
\usepackage[mode=buildnew]{standalone} 
\usepackage{tikz-cd}
\usepackage{tikz}
\usepackage[dvipsnames]{xcolor}
\usepackage{pgfplots}
\usepgfplotslibrary{fillbetween}
\pgfplotsset{compat=1.18}
\usetikzlibrary{automata, arrows, positioning, calc, fit, arrows.meta, shapes, patterns, decorations.markings, backgrounds}

\newcommand{\re}[1]{\textcolor{black}{#1}}
\newcommand{\asinh}[0]{\text{asinh} \,}
\newcommand{\sigmaopt}[0]{\sigma^*}
\newcommand{\sigmacons}[0]{\sigma^\text{cons}}

\newcounter{appcount}
\renewcommand{\theappcount}{Appendix \Alph{appcount}}

\makeatletter
\newcommand{\appsection}[1]{%
  \refstepcounter{appcount}%
  \par
  \vspace{.5ex \@plus .5ex \@minus .2ex}
  \indent
  {\normalfont\normalsize\itshape \theappcount: #1}
  \@afterindenttrue 
}
\makeatother

\begin{document}

\preprint{APS/123-QED}

\title{Near-optimality of conservative driving in discrete systems
}%

\author{Jann van der Meer}
\email{vandermeer.jann.57d@st.kyoto-u.ac.jp}
\author{Andreas Dechant}%
\email{andreas.dechant@outlook.com}
\affiliation{%
Department of Physics No. 1, Graduate School of Science, Kyoto University, Kyoto 606-8502, Japan
}%

\date{\today}

\begin{abstract}
Transferring a physical system from an initial to a final state while minimizing energetic losses is an interdisciplinary control problem that bridges stochastic thermodynamics and optimal transport theory. 
Recent research typically considers problems in which the optimal solution is realized via conservative forces, but whether this situation applies depends on the problem's constraints. 
In systems with complex topologies like discrete networks, the optimal, dissipation-minimizing protocol involves applying nonconservative forces along cycles if the timescales of the transitions in the network are fixed. 
We show that although nonconservative driving is optimal in this setting, a conservative protocol \re{always} exists whose dissipation \re{is at most 4/3 times the optimum}.
This finding is complemented with an example modeling transport across an energy barrier, which \re{saturates this bound}. 
\re{We also prove near-optimality of conservative driving with a weaker numerical factor for a more general class of rate parametrizations with different load-sharing factors, stimulating the idea that} optimality of nonconservative driving might be a generic phenomenon:  
As fewer degrees of freedom can be optimized, additional degrees of freedom due to adding nonconservative forces become more significant.
\end{abstract}

\maketitle

\paragraph{Introduction---}

Optimizing physical systems to minimize their dissipation is of practical importance for the design of energy-efficient devices.
It is therefore crucial to understand the mechanisms that cause dissipation and to discover general strategies to reduce it.
One approach to tackling these issues is the framework of optimal transport \cite{mong81, bena00, vill08}, whose goal is to achieve a desired outcome while minimizing a cost function.
When the cost function is taken to be the entropy production of a physical process, the result are optimal physical forces that allow realizing the process with minimal dissipation \cite{aure11,aure12, naka21,vu23a,oika25}.
While the concrete optimal forces can often only be computed approximately, in many cases the optimal forces can be shown to be conservative \cite{aure11,dech22c,vu23a}, which reduces the optimization problem to the construction of an optimal potential function.

From a physical point of view, nonconservative forces prevent a system from relaxing to an equilibrium state and are themselves a source of dissipation \cite{schn76a, hill89, hata01,maes07,dech22,yosh23}.
However, from the point of view of optimization, conservative forces are merely a tiny subset of all possible forces, so it is surprising that optimizing over the much larger space of nonconservative forces does not improve the result.
Indeed, nonconservative forces enable new phenomena or improve performance in various artificial and biological systems, such as anomalous relaxation \cite{degu22, dieb23}, autonomous feedback \cite{stra13}, coherent oscillations \cite{bara17, ohga23}, cooling \cite{loos20}, collective phenomena in active matter \cite{knez22, fodo22}, enhanced sensing \cite{dech25}, sensory adaptation \cite{lan12}, signal transduction \cite{ito15} and optimal driving in viscous media \cite{loos24}.

One exception to the optimality of conservative forces was discovered in Ref.~\cite{reml21}: In discrete jump processes, the optimal forces minimizing the entropy production rate are generally nonconservative.
This result can be reconciled with other findings \cite{dech22c,ilke22,yosh23,vu23a,naga25} by noting that, for jump processes, minimizing dissipation leads to a non-trivial solution only under additional constraints \cite{mura13,dech22c}.
These constraints represent the physical reality that transitions between different states cannot occur arbitrarily fast; whether the optimal forces are conservative or not depends on the precise constraints.
Constraints such as on the magnitude of transition rates or forces are also ubiquitous in practical applications, where they may prohibit us from realizing the solution prescribed by optimal transport theory.
Even though the latter may be conservative, the optimal forces under the relevant physical constraints may be nonconservative.

This raises two key questions:
First, how can we understand the conditions under which the optimal forces are nonconservative and how can they reduce dissipation compared to conservative forces? 
Second, how much can dissipation be reduced by nonconservative forces and are conservative forces still near-optimal?

In this Letter, we address these questions on the background of Refs.~\cite{mura13, reml21}.
We consider a Markov jump process on a discrete state space, and minimize the entropy production for a given initial and final state, under the constraint that the symmetric part of the transition rates is fixed.
\re{We show that under these conditions conservative forces are near-optimal in the following sense: There is always a conservative protocol whose entropy production exceeds the nonconservative optimum by at most a third. The bound is saturated in a one-dimensional model consisting of a ring with a single large energy barrier.
Physically, optimality of nonconservative forces is due to their greater flexibility in balancing probability flow across the barrier and through the bulk of the ring.
For general load-sharing factors, we prove a structurally identical bound with a weaker, parametrization-dependent constant.}

\paragraph{Minimal dissipation in discrete systems---}

For Markovian dynamics on a set of $N$ discrete states, the probability of finding the system in state $i$, $p_i(t)$, evolves according to the master equation
\begin{equation}
    - \dot{p}_i(t) = \sum_{j \neq i} j_{ij}(t) = \sum_{j \neq i} [ p_i(t) k_{ij}(t) - p_j(t) k_{ji}(t)]
    \label{eq:intro:master}
,\end{equation}
a continuity equation expressing that a change of probability in state $i$ must be due to a current $j_{ij}(t)$ to a neighboring state $j$. The current can also be expressed in terms of $p_i(t)$ and $k_{ij}(t)$, the rate of transitions to state $j$ if the system is in state $i$ at time $t$. From the perspective of optimal transport problems, the master equation \eqref{eq:intro:master} encodes constraints on every possible transport plan imposed by the system's topological features. 

In addition to fixing the topology, the transport problem of moving probability between two states requires specifying the cost function to be optimized. In the context of thermodynamics, a natural and physically well-motivated cost term is given by the total entropy production $S = \int_0^T \sigma dt$ between two specified times $0$ and $T$. The integrand
\begin{equation}
    \sigma =  \sum_{i, j} p_i(t) k_{ij}(t) \ln \frac{p_i(t) k_{ij}(t)}{p_j(t) k_{ji}(t)}
    \label{eq:intro:sigma}
\end{equation}
is the instantaneous entropy production rate in units of Boltzmann's constant $k_B = 1$ \cite{schn76a, seif25}. Without any additional constraints, infinitely fast but almost symmetric transitions allow transporting probability with vanishing entropy production, thereby trivializing the optimization problem \cite{mura13}. We can fix the timescale of the transitions in the system by parameterizing the transition rates as \cite{kolo07, mura13, reml21, seif25}
\re{\begin{equation}
    k_{ij}(t) = \kappa_{ij}(t) e^{\alpha_{ij}(t) A_{ij}(t)}
    \label{eq:intro:k_param_alpha}
\end{equation}
with a nonnegative symmetric matrix $\kappa_{ij}(t) = \kappa_{ji}(t) \geq 0$, an antisymmetric matrix $A_{ij}(t) = -A_{ji}(t)$, and load-sharing (or load-distribution \cite{fish99, kolo07}) factors $\alpha_{ij}(t) = 1 - \alpha_{ji}(t)$, $0 \leq \alpha_{ij}(t) \leq 1$.} This parametrization has a clear thermodynamic interpretation: $A_{ij}(t)$ models the forces in the system, whereas the symmetric part encodes kinetic information, e.g., the presence of energy barriers between states and the connectivity of the state space. We say that two states $i$ and $j$ are connected when $\kappa_{ij}(t) > 0$, which we interpret as an edge in the graph whose vertices are the states $i$. \re{The load-sharing factors allow asymmetric splittings of the nonequilibrium forces onto forward and backward rates due to asymmetric energy landscapes, a crucial feature for, e.g., molecular motor models \cite{fish99, kolo07, zimm12, erte23}. Assuming control over only the forces $A_{ij}(t)$ in the system at fixed $\kappa_{ij}(t)$ and $\alpha_{ij}(t)$ results in a nontrivial optimization problem for $0 < \alpha_{ij}(t) < 1$. For $\alpha_{ij}(t) = 1/2$, Eq. \eqref{eq:intro:k_param_alpha} simplifies to
\begin{equation}
    k_{ij}(t) = \kappa_{ij}(t) e^{A_{ij}(t)/2}
    \label{eq:intro:k_param}
\end{equation}
and} the entropy production rate takes the form
\begin{equation}
    \sigma = \sum_{i,j} \omega_{ij}(t) F_{ij}(t) \sinh \frac{F_{ij}(t)}{2}
    \label{eq:intro:sigma_2}
.\end{equation}
The symmetric matrix $\omega_{ij}(t) := \kappa_{ij}(t) \sqrt{p_i(t) p_j(t)}$ sets the timescale of transitions between states $i$ and $j$.
The expression \eqref{eq:intro:sigma_2} is convex in the thermodynamic forces
\begin{align}
    F_{ij}(t) := A_{ij}(t) + \ln (p_i(t)/p_j(t)) = 2 \, \asinh \frac{j_{ij}(t)}{2 \omega_{ij}(t)}
    \label{eq:intro:def_F}
,\end{align}
and the system is in equilibrium if and only if $F_{ij}(t) = 0$ for every pair of states for which $\omega_{ij}(t) > 0$. \re{For an equilibrium system, $\omega_{ij}(t)$ coincides with the traffic.}
We say that the forces in the system are conservative if a state function $\psi_i(t)$ exists such that $F_{ij}(t) = \psi_i(t) - \psi_j(t)$, which is equivalent to $A_{ij}(t) = \phi_i(t) - \phi_j(t)$ with $\phi_i(t) = \psi_i(t) - \ln (p_i(t))$.

Being able to change $A_{ij}(t)$ is equivalent to being able to control $F_{ij}(t)$ or the current $j_{ij}(t)$ at each transition.
Optimizing over $F_{ij}(t)$ ($M$ degrees of freedom (DOF) if the graph has $M$ edges) can be reparametrized and equivalently understood as an optimization over all $\dot{p}_i(t)$ ($N-1$ DOF for an $N$-state system), followed by a constrained optimization over all cycle currents ($C = M - (N - 1)$ DOF for a system with $C$ fundamental cycles). This latter optimization over currents $j_\mathcal{C}$ along a cycle $\mathcal{C}$ corresponds to adjusting circular flows in the network and can be performed for fixed $\dot{p}_i(t)$, i.e., without affecting the time-evolution of the system, \re{in accordance with Kirchhoff's current law \cite{schn76a, hill89, bond07}. Operationally, we can take} the derivative with respect to a cycle current $\partial_{j_\mathcal{C}}$ as explained in \ref{appendix:cycles}.

Therefore, and because entropy production is \re{time-additive} in a Markovian system, the optimization over the cycle currents $j_\mathcal{C}$ can be performed independently for each point in time $t$. 
Thus, we can consider a fixed point in time $t$ and suppress the time arguments. Minimizing the entropy production rate then corresponds to the condition
\begin{equation}
    \partial_{j_\mathcal{C}} \sigma = 2 \sum_{(ij) \in \mathcal{C}} \left( \tanh (F_{ij}/2) + F_{ij}/2 \right) = 0 .
    \label{eq:intro:dsigma_djc}
\end{equation}
On the other hand, for conservative forces $F_{ij} = \psi_i - \psi_j$, the cycle affinity
\begin{equation}
    \mathcal{A}_\mathcal{C} = \sum_{(ij) \in \mathcal{C}} F_{ij} = \sum_{(ij) \in \mathcal{C}} A_{ij}
    \label{eq:intro:affinity}
\end{equation}
vanishes for any cycle $\mathcal{C}$.
Comparing \eqref{eq:intro:dsigma_djc} and \eqref{eq:intro:affinity}, we see that, surprisingly, the conditions for minimal entropy production and vanishing cycle affinity are different and, thus, optimal driving generally requires nonconservative forces \re{\cite{reml21}}.

\paragraph{Conflicting intuitions---}

\begin{figure}[t]
    \centering
    \includestandalone[scale=1]{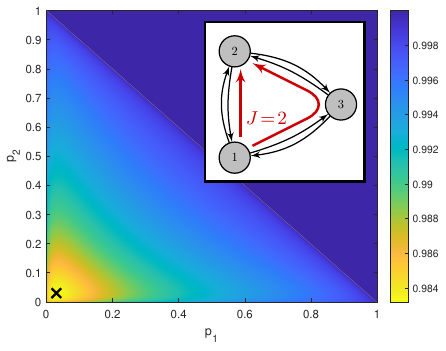}
    \caption{Optimal versus conservative driving in a three-state Markov network. Colors encode the maximal possible improvement $\sigma^*/\sigma^{\text{cons.}}$ according to the color scale. For given occupation probabilities $p_i$, the configurations that yield a desired current $J = \dot{p}_2 = - \dot{p}_1$ can be parametrized by the cycle affinity $\mathcal{A} = A_{12}+A_{23}+A_{31}$. The minimal thermodynamic cost \eqref{eq:intro:sigma} $\sigma^*$ after optimizing $\mathcal{A}$ is compared to the cost $\sigma^{\text{cons.}}$ of the conservative protocol satisfying $A_{12}+A_{23}+A_{31} = 0$. The transition rates are parametrized as in Eq. \eqref{eq:intro:k_param} with fixed ${\kappa_{ij} = 1}$ for $i\neq j$. For $J = 2$, the minimum $\sigma^*/\sigma^{\text{cons.}} = 0.983$ is located at $(p_1, p_2) = (0.03, 0.03)$.}
    \label{fig:fig0}
\end{figure}

This perhaps surprising insight leaves us with at first glance contradictory perspectives. 
On the one hand, nonconservative driving generates cycle currents that generally increase dissipation, which prompts the question how nonconservative driving can be optimal. \re{In particular, in steady states all dissipation is due to nonconservative forces.}
On the other hand, the same cycle currents correspond to additional optimizable degrees of freedom enabled by nonconservative driving forces, in turn provoking the question why conservative driving is optimal in many previously studied situations, e.g., for constant activity \cite{dech22c, naga25}, in the continuum limit \cite{reml21, dech22c} \re{and for slow driving \cite{ilke22}, particularly in the context of thermodynamic friction geometry \cite{mand16, sawc26}}. 
How can we reconcile these seemingly contradictory viewpoints? More practically, how do we find situations in which additional nonconservative forces offer significant improvement?

A first insight is that discrepancies between Eqs. \eqref{eq:intro:dsigma_djc} and \eqref{eq:intro:affinity} are only observed at order $\mathcal{O}(F_{ij}^3)$, implying that the optimal and conservative protocols coincide near equilibrium. 
\re{However, making $F_{ij}$ large by taking $p_j\to0$ is not a useful far-from-equilibrium limit: At fixed nonzero probability flux at state $j$, $\sigma\geq |\dot p_j|\ln(1/p_j)+O(1)$, and cycle currents can only increase the singular prefactor. Thus, optimizing nonconservative forces cannot reduce the singular part of the cost in this limit \cite{shortpaper_SM}.}
In a numerical case study \cite{reml21}, the improvement of optimal nonconservative over conservative protocols has been reported to be of a relative magnitude of $\sim10^{-2}$, which is also confirmed by optimizing over the possible probabilities $p_i$ in a three-state system in Fig.\,\ref{fig:fig0}. Even when focusing on the nonadiabatic entropy production \cite{espo10, seif25}, i.e., when disregarding the cost of the stationary cycle current, a numerical case study shows limited improvements that vanish in both limits of slow and fast driving \cite{delv24}. 
What is lacking is a way to systematically access and bound the difference between the costs of the conservative and optimal protocol. This is the purpose of the following \re{two main results}.

\paragraph{\re{Main result I: General load-sharing factors---}}

\begin{figure*}[t]
    \centering
    \includestandalone[scale=1]{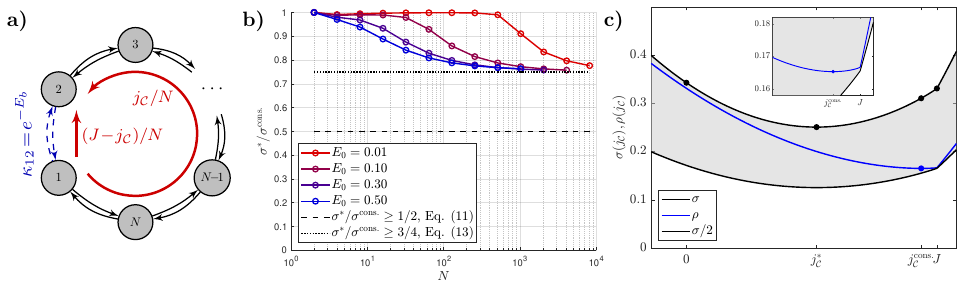}
    \caption{\re{Transport across an energy barrier in a unicyclic network. a) Set-up. We take $p_i=1/N$, $\kappa_{12}=e^{-E_b}$ with an energy barrier $E_b=NE_0$, and $\kappa_{ij}=1$ otherwise for connected $i,j$.
    We impose the probability flux $\dot p_2=-\dot p_1=J/N$; the bulk edges transport $j_{\mathcal C}/N$, whereas the barrier edge transports $j_{12}=(J-j_{\mathcal C})/N$.   
    b) Comparison of conservative and optimal transport costs $\sigmacons$ and $\sigmaopt$. For every $N, E_0, J$, $\sigma^*(J)$ is obtained by minimizing
    $\sigma(j_{\mathcal C};J)$ (cf. Eq.~\eqref{eq:num:sigma}) over $0\leq j_{\mathcal C}\leq J$, whereas $\sigmacons (J)$ is evaluated at zero cycle affinity.
    Each point uses the $J$ numerically minimizing $\sigmaopt (J)/\sigmacons (J)$. The optimized ratio approaches $3/4$  for large $N$.
    c) Comparison of $\rho(j_{\mathcal C}) := \rho^{1/2}(j_{\mathcal C})$ (cf. Eq. \eqref{eq:mainres:sigma}) to $\sigma(j_\mathcal{C})$ and $\sigma(j_{\mathcal C})/2$. 
    For the depicted parameters $N=100$, $E_0 = 0.3$, $J=0.58$ minimizes $\sigmaopt (J)/\sigmacons (J)$. 
    The minima of $\sigma$ and $\rho$ lie at $j_{\mathcal C}^*=0.30$ and $j_{\mathcal C}^{\mathrm{cons.}}=0.55$, respectively.}}
    \label{fig:fig1}
\end{figure*}

\re{Our first main result applies to the general rate parametrization Eq. \eqref{eq:intro:k_param_alpha}. Introducing $\omega_{ij}^\alpha := \kappa_{ij} p_i^{1 - \alpha_{ij}} p_j^{\alpha_{ij}}$ and, similar to Eq. \eqref{eq:intro:def_F}, $F_{ij} = A_{ij} + \ln (p_i/p_j)$ we have
\begin{align}
    \sigma = & \, \frac{1}{2} \sum_{ij} \omega_{ij}^\alpha F_{ij} \left( e^{\alpha_{ij} F_{ij}} - e^{- (1 - \alpha_{ij}) F_{ij}} \right) \label{eq:mainres:sigma} \\
    \rho^\alpha := & \, \sigma - \frac{1}{2} \sum_{ij}  \omega_{ij}^\alpha \left(  \frac{e^{\alpha_{ij} F_{ij}} - 1}{\alpha_{ij}} + \frac{e^{- (1 - \alpha_{ij}) F_{ij}} - 1}{1 - \alpha_{ij}}\right) \nonumber
.\end{align}
The first line generalizes Eq. \eqref{eq:intro:sigma_2} to general $\alpha_{ij}$, whereas the second line introduces a different measure of distance to equilibrium $\rho^\alpha$ with two crucial properties. First, it satisfies
\begin{equation}
    \sqrt{\alpha(1-\alpha)} \sigma \leq \rho^\alpha \leq \sigma
    \label{eq:mainres:rho_sigma_bounds}
\end{equation}
for $\alpha = \min_{i, j} \alpha_{ij}$, as proved in the SM \cite{shortpaper_SM}. Second, conservative forces minimize $\rho^\alpha$. For example, an explicit calculation  \cite{shortpaper_SM} gives $\partial \rho^{1/2}/\partial j_\mathcal{C} = \mathcal{A}_\mathcal{C}$ in the case $\alpha = 1/2$.
A brief thermodynamic interpretation of $\rho^\alpha$ and the derivation of Eq. \eqref{eq:mainres:rho_sigma_bounds} are given in \ref{appendix:load_sharing}. We explore additional thermodynamic properties in the case $\alpha = 1/2$ in Ref. \cite{longpaper}.}

\re{Comparing $\sigma$ to $\rho^\alpha$ allows us to quantitatively relate optimal and conservative transport.} At each point in time $t$, the minimal entropy production rate $\sigmaopt$ associated with the optimal protocol bounds the entropy production rate $\sigmacons$ of the corresponding conservative protocol with the same time evolution from above and below,
\begin{equation}
    \sigmaopt \leq \sigmacons \leq \frac{1}{\sqrt{\alpha(1-\alpha)}} \sigmaopt
    \label{eq:mainres:bounds}
,\end{equation}
which can be derived from Eq. \eqref{eq:mainres:rho_sigma_bounds} as shown in \ref{appendix:bounds}. By integrating over the duration $T$ of a protocol, we obtain bounds on the total entropy production $S^* := \int_0^T \sigmaopt dt $ of the optimal protocol\re{
\begin{equation}
    \sqrt{\alpha_{\mathrm{min}}(1-\alpha_{\mathrm{min}})} S^\text{cons} \leq S^* \leq S^\text{cons} := \int_0^T \sigmacons dt
    \label{eq:mainres:bounds_integrated}
,\end{equation}
with $\alpha_{\mathrm{min}} := \min_t \alpha(t)$}. This relation establishes that conservative protocols, while not necessarily optimal, are near-optimal in the sense that deviations from the optimal solution are bounded by the constant numerical factor of \re{$1/\!\sqrt{\alpha_{\mathrm{min}}(1-\alpha_{\mathrm{min}})}$}. \re{We note that the optimal conservative protocol might outperform $S^\text{cons}$; this optimization problem is discussed for $\alpha = 1/2$ in Ref. \cite{longpaper}}.

\paragraph{\re{Main result II: Tight bound for $\alpha_{ij} = 1/2$---}}
In this case, the higher symmetries of parametrization \eqref{eq:intro:k_param} allow us to prove the tighter bound
\begin{equation}
    \sigmaopt \leq \sigmacons \leq 4 \sigmaopt/3
    \label{eq:mainres2:bounds}
,\end{equation}
which means that, given an optimal transport problem, conservative forces provide a sensible trial solution whose cost can be reduced by at most the factor $1/4$, as contrasted to $1/2$ when applying Eq. \eqref{eq:mainres:bounds_integrated}. We derive the bound in \ref{appendix:proof_mainres2} and proceed with an example illustrating that a single, high energy barrier in a unicyclic topology suffices to saturate the factor $4/3$.

We \re{consider an infinitesimal time step and} assume a uniform probability distribution $p_i = 1/N$ over $N$ states arranged in a cycle. 
\re{We set $\kappa_{12}=e^{-E_b}$ and
$\kappa_{ij}=1$ on all other nearest-neighbor edges,
modeling an energy barrier $E_b > 0$ between states $1$ and $2$, cf. Fig.\,\ref{fig:fig1}\,a). 
In Fig.\,\ref{fig:fig0}, the optimal solution features a slower timescale $\omega_{12}$ of the edge directly connecting source and target as compared to the other edges, numerically $\omega_{12} \simeq 0.17 \, \omega_{23} = 0.17 \, \omega_{31}$. Adjusting the energy barrier $E_b > 0$ allows tuning this timescale separately without modifying the occupation probabilities $p_i$.}

Transporting probability at the rate $J/N = \dot{p}_2 = - \dot{p}_1$ from state $1$ to $2$ can be achieved either through a local force $A_{12}$ that directly affects the current $j_{12}$ across the energy barrier or by transporting probability through the bulk of the cycle, which we parametrize as $j_\mathcal{C}/N = J/N - j_{12}$. Using Eqs. \eqref{eq:intro:sigma_2} and \eqref{eq:intro:def_F}, the entropy production rate  
\begin{equation}
    \sigma = \frac{2(N - 1)}{N} j_{\mathcal{C}} \, \asinh \frac{j_{\mathcal{C}}}{2} + \frac{2}{N} (J - j_{\mathcal{C}}) \, \asinh \frac{J - j_{\mathcal{C}}}{2 e^{-E_b}}
    \label{eq:num:sigma}
\end{equation}
can be optimized as a function of $j_\mathcal{C}$. The minimizer $j_{\mathcal{C}}^*$ can be contrasted with $j_\mathcal{C}^{\text{cons.}}$, the value of $j_{\mathcal{C}}$ that solves $0 = 2 (N - 1) \asinh (j_{\mathcal{C}}/2) - 2 \, \asinh ([J - j_{\mathcal{C}}]/[2 e^{-E_b}])$ and therefore indicates conservative driving (\re{see SM~\cite{shortpaper_SM} for details}). For large $N$ and suitably scaled $E_b = N E_0$ with constant $E_0$, the solutions $j_{\mathcal{C}}^*$ and $j_\mathcal{C}^{\text{cons.}}$ show different qualitative behavior. 
\re{Numerically, we obtain up to $\sigma^*/\sigma^{\text{cons.}} \simeq 0.76$, as depicted in Fig.\,\ref{fig:fig1}\,b). In the SM~\cite{shortpaper_SM}, we prove that $\sigma^*/\sigma^{\text{cons.}} \to 3/4$ in the limit of first $N \to \infty$, then $J = 2 E_0 \to 0$.} Qualitatively, an explicit illustration of Eq. \eqref{eq:num:sigma} in Fig.\,\ref{fig:fig1}\,c) reveals that the optimal current $j_{\mathcal{C}}^*$ balances transport through the bulk with transport across the barrier, whereas the larger $j_\mathcal{C}^{\text{cons.}}$ indicates that transport through the barrier is mostly avoided. \re{This effect becomes more pronounced for larger $N$, as illustrated in \re{the SM~\cite{shortpaper_SM}}, and is also observed in numerical optimizations of finite-time protocols~\cite{longpaper}}.

\paragraph{Discussion and outlook---} 
Model classes in which nonconservative driving is optimal are less explored than scenarios in which conservative driving can achieve optimality. Our results show that properties of the optimal solution depend crucially on a problem's constraints, e.g., which parameters are fixed and which can be optimized. It is expected that the fewer DOF available for optimization, the more significant the additional DOF gained through nonconservative driving become. This suggests that, rather than being an exception, optimality of nonconservative driving may be the generic case in real-world scenarios involving manifold and complex constraints. \re{For example, our results indicate that conservative solutions are less optimal in the presence of strongly asymmetric load-sharing factors.}

A concrete comparison can be drawn \re{with Ref. \cite{dech22c}, which optimizes entropy production for different constraints}: 
\re{In Ref. \cite{dech22c}, only the total activity is fixed, so activity can be redistributed among individual edges. This looser condition permits the additional freedom to ``turn off'' individual edges. Here, by contrast, every kinetic prefactor
$\kappa_{ij}$ is fixed individually. In a nontrivial topology all pathways remain relevant, so that a nontrivial optimization over cycle-current degrees of freedom must be performed. In this situation, nonconservative forces offer greater flexibility than conservative forces alone.}

The mutable nature of optimal solutions depending on their constraints stimulates a shift in attention in problems of minimal cost towards the constraints themselves. 
\re{The parametrization of rates adopted in this work is not exhaustive; one can, e.g., consider general nonlinear relationships between control parameters and transition rates.}
To explore this vast realm of possible system classes, it is also worthwhile to investigate the microscopic origin of possible rate parameterizations. A theoretical starting point is to coarse-grain a Langevin model of, say, a bistable or tilted potential in the presence of a control parameter for the driving, and then investigate the rate parameterizations emerging in an effective discrete Markovian approximation. \re{Another question of interest is whether the configuration of cycle currents that instantaneously minimizes the entropy production rate admits a decomposition into housekeeping and excess contributions to entropy production similar to Refs. \cite{maes14, dech22}. Finally, investigating potential tighter bounds improving on $\sigma^*/\sigma^{\text{cons.}} \geq \sqrt{\alpha(1-\alpha)}$ seems promising given the existence of a tighter, saturable bound in the case $\alpha = 1/2$. We leave this as an open problem for future work.}

\paragraph{Acknowledgments---} We thank Masato Itami for insightful discussions.
JvdM was supported by JSPS KAKENHI (Grants No. 24H00833 and 25KF0238) and JSPS Research Fellowship No. P25761.
AD was supported by JSPS KAKENHI (Grants No. 22K13974, 24H00833 and 25K00926).
This work was supported by JST ERATO Grant Number JPMJER2302, Japan.

\bibliography{references}

\pagebreak
\clearpage

\section{End matter}
\appendix

\begin{figure}[ht!]
    \centering
    \includestandalone[scale=1]{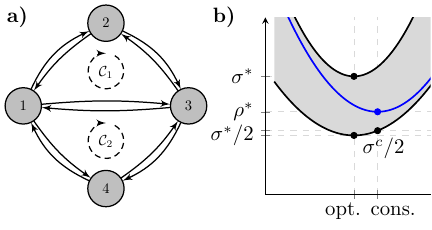}
    \caption{a) A Markov network with two fundamental cycles $\mathcal{C}_1 = (1231)$ and $\mathcal{C}_2 = (1341)$. 
    In this example, $\partial_{j_{\mathcal{C}_1}} = \partial_{j_{12}} + \partial_{j_{23}} - \partial_{j_{13}}$. 
    Note that increasing the edge currents $j_{12}, j_{23}, j_{31} = - j_{13}$ by the same amount does not affect a given time evolution of the occupation probabilities $p_i(t)$.
    b) \re{Visualization of the derivation that a relation $\sigma/K \leq \rho \leq \sigma$ between two cost functions $\sigma$, $\rho$ implies the same chain of inequalities on their minimal values $\sigma^*$, $\rho^*$, as illustrated for $K = 2$. We can relate $\sigma^c$, the value of $\sigma$ evaluated at the point at which $\rho$ is minimized, to its minimum $\sigma^*$ via $\sigma^c \leq K \sigma^*$.} In the main text, the configurations minimizing $\sigma$ and $\rho$ characterize optimal and conservative driving, respectively.}
    \label{fig:fig2}
\end{figure}

\appsection{Cycles and cycle currents---} 
\label{appendix:cycles}
The time evolution of the probabilities $p_i(t)$ for all times $t$ according to Eq. \eqref{eq:intro:master} does not determine the time evolution of the currents $j_{ij}(t)$ from state $i$ to $j$ if the system's topology is nontrivial and contains cycles. 
The remaining degrees of freedom are specified by the cycle currents on a set of fundamental cycles \cite{schn76a}, which form a basis for all possible cycles \re{in the steady state \cite{schn76a, hill89} or when the changes in probability $\dot{p}_i(t)$ are fixed \cite{bond07}, see} Fig.\,\ref{fig:fig2}\,a) for an example. 
We understand a cycle $\mathcal{C}$ as a path (a sequence of directed edges) in the network that enters a state at most once and whose first and last state coincide, forming a closed loop. We write $(ij) \in \mathcal{C}$ if the transition $i \to j$ is contained in $\mathcal{C}$.
The directed adjacency matrix of a cycle $\mathcal{C}$ is defined as
\begin{align}
   \chi^{\mathcal{C}}_{ij} = \begin{cases}
        1 & \text{ if } i \to j \text{ is contained in } \mathcal{C} \\
        - 1 & \text{ if } j \to i \text{ is contained in } \mathcal{C} \\
        0 & \text{ otherwise }
    \end{cases}
,\end{align}
which enables us to rigorously define the derivative with respect to the cycle current $\partial_{j_\mathcal{C}}$ as  
\begin{align}
    \partial_{j_\mathcal{C}} = \sum_{(ij) \in \mathcal{C}} \partial_{j_{ij}} = \sum_{i < j} \chi^{\mathcal{C}}_{ij} \partial_{j_{ij}} = \frac{1}{2} \sum_{ij} \chi^{\mathcal{C}}_{ij} \partial_{j_{ij}}
    \label{eq:app_a:def_djc}
,\end{align}
where $i<j$ ensures that each edge is contained exactly once in the sum. The decomposition of cycles into fundamental cycles (e.g. $\mathcal{C} = \mathcal{C}_1 + \mathcal{C}_2$) translates into a corresponding linear combination of differential operators (e.g. $\partial_{j_\mathcal{C}} = \partial_{j_{\mathcal{C}_1}} + \partial_{j_{\mathcal{C}_2}}$.

\appsection{Bounds on cost functions imply bounds on their minima---} 
\label{appendix:bounds}
We consider two cost functions $\sigma(x), \rho(x)$ on the same domain satisfying 
\begin{equation}
    \sigma/K \leq \rho \leq \sigma
    \label{eq:app_b:dassump}
\end{equation}
\re{for a constant $K \geq 1$} and denote their respective minima as $\sigma^*$ and $\rho^*$. For $x_c$, defined as the $x$-value minimizing $\rho$, we have $\sigma^* \leq \sigma^c := \sigma(x_c) \leq K \rho^*$ by virtue of the first inequality in \eqref{eq:app_b:dassump}. For $x_o$, the $x$-value that minimizes $\sigma$, we have $\rho^* \leq \rho(x_o) \leq \sigma(x_o) = \sigma^*$, where the second inequality makes use of \eqref{eq:app_b:dassump}. Combining these two relations, we obtain
\begin{align}
    \sigma^* \leq \sigma^c \leq K \rho^* \leq K \sigma^*
,\end{align}
which includes the finding $\sigma^c \leq K \sigma^*$ that is crucial to our main results. The schematic idea of the proof is illustrated in Figure \ref{fig:fig2}\,b). We can also conclude that the minimum of one of the functions $\sigma$, $\rho$ also yields a strong, quantitative constraint on the respective other minimum, e.g., \re{$\rho^* \le \sigma^* \le K \rho^*$}, meaning that the minimum of $\sigma$ is bounded from below and above by the minimum of $\rho$.

\appsection{\re{Derivation and interpretation of $\rho^\alpha$ for asymmetric load-sharing factors}---} 
\label{appendix:load_sharing}
\re{In this section, we sketch the proof of Eq. \eqref{eq:mainres:bounds} for the rate parametrization given as Eq. \eqref{eq:intro:k_param_alpha},
\begin{equation}
    k_{ij}(t) = \kappa_{ij}(t) e^{\alpha_{ij}(t) A_{ij}(t)}, \, 0 \leq \alpha_{ij}(t) = 1 - \alpha_{ji}(t) \leq 1 \nonumber
,\end{equation}
with symmetric $\kappa_{ij}(t) = \kappa_{ji}(t)$ and antisymmetric $A_{ij}(t) = - A_{ji}(t)$.
The current $j_{ij}(t) = p_i(t) k_{ij}(t) -  p_j(t) k_{ji}(t)$ can be expressed as 
\begin{align}
    j_{ij}(t) & = \kappa_{ij}(t) \left( p_i(t) e^{\alpha_{ij}(t) A_{ij}(t)} - p_j(t) e^{\alpha_{ji}(t) A_{ji}(t)} \right) \nonumber \\
    & = \omega_{ij}^\alpha(t) \left( e^{\alpha_{ij}(t) F_{ij}(t)} - e^{- [1 - \alpha_{ij}(t)] F_{ij}(t)} \right)
    \label{eq:app_d:current_f}
\end{align}
by introducing $\omega_{ij}^\alpha(t) := \kappa_{ij}(t) p_i(t)^{1 - \alpha_{ij}(t)} p_j(t)^{\alpha_{ij}(t)}$ and $F_{ij}(t) = A_{ij}(t) + \ln (p_i(t)/p_j(t))$ as in Eq. \eqref{eq:mainres:sigma}. The entropy production rate retains the familiar form $\sigma = (1/2) \sum_{ij} j_{ij}(t) F_{ij}(t) = \sum_{i < j} j_{ij}(t) F_{ij}(t)$. Dropping the time argument for ease of notation, we calculate the derivative of the current $j_{ij}$ with respect to $F_{ij}$,
\begin{align}
    \frac{\partial j_{ij}}{\partial F_{ij}} = \omega_{ij}^\alpha \left( \alpha_{ij} e^{\alpha_{ij} F_{ij}} + [1 - \alpha_{ij}] e^{- [1 - \alpha_{ij}] F_{ij}} \right)
    \label{eq:app_d:djdf}
,\end{align}
whose inverse enters the calculation
\begin{align*}
    \frac{\partial \sigma}{\partial j_{ij}} & = \frac{\partial}{\partial j_{ij}} \left( \sum_{i < j} j_{ij} F_{ij}(j_{ij}) \right)  = F_{ij} + j_{ij} \left( \frac{\partial j_{ij}}{\partial F_{ij}} \right)^{-1} \\
    & = F_{ij} + \frac{e^{\alpha_{ij} F_{ij}} - e^{- [1 - \alpha_{ij}] F_{ij}}}{\alpha_{ij} e^{\alpha_{ij} F_{ij}} + [1 - \alpha_{ij}] e^{- [1 - \alpha_{ij}] F_{ij}}}
.\end{align*}
We see that, similar to the time-symmetric case, the derivative with respect to the cycle current $\partial \sigma/\partial j_\mathcal{C}$ will generally not result in a multiple of the cycle affinity $\mathcal{A}_\mathcal{C} = \sum_{(ij) \in \mathcal{C}} F_{ij}$ but contains an additional nonlinear term. As in the symmetric case, Eq. \eqref{eq:intro:dsigma_djc}, optimality under variations of a cycle current will generally not coincide with the condition of vanishing affinity. This property is a consequence of the nonlinear relationship \eqref{eq:app_d:current_f} between $j_{ij}$ and $F_{ij}$ and, ultimately, of the non-quadratic nature of the cost functional $\sigma$ as a function of either.}

\re{One can, however, redefine the cost functional to a quantity $\rho^\alpha$ and enforce the property $\partial \rho^\alpha/\partial j_{ij} = F_{ij}$. This relation implies $\partial \rho^\alpha/\partial j_{\mathcal{C}} = \mathcal{A}_\mathcal{C} = \sum_{(ij) \in \mathcal{C}} F_{ij}$ for every cycle $\mathcal{C}$. Thus, the global minimum must satisfy $\mathcal{A}_\mathcal{C} = 0$ for every cycle $\mathcal{C}$, meaning that conservative driving minimizes $\rho^\alpha$ (for the case $\alpha_{ij} = 1/2$, cf. Eqs. \eqref{eq:intro:dsigma_djc} and \eqref{eq:intro:affinity}, see the SM \cite{shortpaper_SM} for details). Thus, we define $\rho^\alpha$ as the primitive
\begin{align}
    \rho^\alpha & = \sum_{i < j} \int_0^{j_{ij}} F_{ij}(\widetilde{j}_{ij}) d\widetilde{j}_{ij} \label{eq:app_d:rho_def} \\
    & = \sum_{i < j} \int_0^{F_{ij}(j_{ij})} \widetilde{F}_{ij} \frac{\partial j_{ij}}{\partial F_{ij}}\big( \widetilde{F}_{ij}  \big) \,d \widetilde F_{ij} \nonumber
.\end{align}
To carry out the integration, we use Eq. \eqref{eq:app_d:djdf} and note that $G(x) = (x - 1/\alpha)e^{\alpha x}$ is an antiderivative of $g(x) = \alpha x e^{\alpha x}$. The calculation yields
\begin{align*}
    \rho^\alpha & = \sum_{i < j}  \omega_{ij}^\alpha \int_0^{F_{ij}}\left( \alpha_{ij} e^{\alpha_{ij} x} + [1 - \alpha_{ij}] e^{- [1 - \alpha_{ij}] x} \right) x d x \\
    & = \sum_{i < j}  2 \omega_{ij}^\alpha C^{\alpha_{ij}}(F_{ij}) = \sum_{ij}  \omega_{ij}^\alpha C^{\alpha_{ij}}(F_{ij})
,\end{align*}
where we have defined the function
\begin{align*}
    C^{\alpha_{ij}}(x) := & \, \frac{x}{2} \left( e^{\alpha_{ij} x} - e^{- (1 - \alpha_{ij}) x} \right) \\ & - \frac{1}{2} \left(  \frac{e^{\alpha_{ij} x}}{\alpha_{ij}} + \frac{e^{- (1 - \alpha_{ij}) x}}{1 - \alpha_{ij}} - \frac{1}{\alpha_{ij}(1 - \alpha_{ij})} \right)
.\end{align*}
For $\alpha_{ij} = 1/2$ for all $i,j$, the quantities $\rho^\alpha$ and $C^{\alpha_{ij}}(x)$ reduce to their respective main-text counterparts $\rho$ and $C(x)$.
In the SM, we prove the bounds
\begin{equation}
    \sqrt{\alpha(1-\alpha)} \sigma \leq \rho^\alpha \leq \sigma
\end{equation}
stated as Eq. \eqref{eq:mainres:rho_sigma_bounds} in the main text. The bound \eqref{eq:mainres:bounds},
\begin{equation}
    \sigmaopt \leq \sigmacons \leq \frac{1}{\sqrt{\alpha(1-\alpha)}} \sigmaopt
,\end{equation}
and its counterpart \eqref{eq:mainres:bounds_integrated} for the time-integrated quantities follow from the previous equation by employing the reasoning of \ref{appendix:bounds} for $K = 1/\sqrt{\alpha(1-\alpha)}$.}

\re{The definition of $\rho^\alpha$ as a primitive in Eq. \eqref{eq:app_d:rho_def} suggests comparing its differential to that of $\sigma$, which we write symbolically as $d \rho^\alpha = \sum_{i < j} F_{ij} dj_{ij}$ and $d \sigma = \sum_{i < j} (F_{ij} dj_{ij} + j_{ij} dF_{ij})$, respectively. We see that $\rho^\alpha$ is the unique cost function for which a slight increase in current $j_{ij}$ incurs an additional cost $F_{ij}$ and which vanishes in equilibrium for vanishing $F_{ij}$. In the case of a linear relationship $j_{ij} \propto F_{ij}$, $\rho^\alpha$ is a multiple of $\sigma$, so for both cost functions a variation of a cycle current $j_\mathcal{C}$ incurs a cost proportional to the cycle affinity $\mathcal{A}_\mathcal{C}$.}

\appsection{\re{Proof of main result II}---} 
\label{appendix:proof_mainres2}
\re{We first prove a general bound for odd, increasing, concave functions:
\paragraph{Lemma.}
Let $\ell:\mathbb R\to\mathbb R$ be odd, increasing, concave on
$[0,\infty)$, and satisfy $\ell(0)=0$.  Then, for all $x,y\in\mathbb R$
\begin{equation}
    y\ell(x)
    \leq
    y\ell(y)+\frac14 x\ell(x). \label{eq:app_c:lemma}
\end{equation}}
\re{\paragraph{Proof.} For $xy\leq0$ the left-hand side is nonpositive, whereas both terms on
the right-hand side are nonnegative, so Eq. \eqref{eq:app_c:lemma} is satisfied.
In the case \(xy>0\),  it suffices to consider $x,y>0$ because $\ell$ is odd.  
If $y\geq x$, monotonicity gives $y\ell(x)\leq y\ell(y)$, which is stronger than Eq.~\eqref{eq:app_c:lemma}.} 

\re{In the crucial case $0 < y < x$,
write $y=zx$, with $z\in(0,1)$. Concavity and \(\ell(0)=0\) imply $\ell(y)=\ell(zx)\geq z\ell(x)$.
This implies the first inequality in the chain
\begin{align}
    y\ell(x)-y\ell(y)
    &\leq y(1-z)\ell(x)  \notag\\
    &= z(1-z)x\ell(x) \leq \frac14 x\ell(x),
\end{align}
whereas the second one follows from $\sup_{z\in(0,1)} z(1-z) = 1/4$ and concludes the proof. $\square$}

\re{To prove Eq. \eqref{eq:mainres2:bounds}, we assume a given time-evolution $\dot{p}_i$ and refer to the unique configuration of currents that minimizes $\sigma$ as $j_{ij}^*$, whereas the currents $j_{ij}^\text{cons}$ are the ones realized with conservative forces. The corresponding thermodynamic forces $F_{ij}^*$ and $F_{ij}^\text{cons}$ can be calculated from Eq. \eqref{eq:intro:def_F}. Whereas $j_{ij}^*$ minimizes $\sigma = \sum_{i<j} j_{ij} F_{ij}$ for fixed $\dot{p}_i$, the current $j_{ij}^\text{cons}$ has the characterization
\begin{align}
    \sum_{i<j} j_{ij}^\text{cons} F_{ij}^\text{cons} = \sum_{i<j} j_{ij} F_{ij}^\text{cons} 
    \label{eq:app_c:cons_char}
\end{align}
for all currents $j_{ij}$ with the correct time-evolution $\dot{p}_i$. Eq. \eqref{eq:app_c:cons_char} follows from conservativity, $0 = \sum_{(ij) \in \mathcal{C}} F_{ij}^\text{cons}$ for all $\mathcal{C}$ (cf. Eq. \eqref{eq:intro:affinity}), by noting that $j_{ij}^\text{cons} - j_{ij}$ is necessarily a linear combination of cycle currents \cite{schn76a, hill89, bond07}.}

\re{We now apply the lemma edgewise to $F_{ij}(j_{ij}) = 2 \, \asinh [j_{ij}/(2 \omega_{ij})]$ (cf. Eq. \eqref{eq:intro:def_F}), with $x = j_{ij}^\text{cons}$, $y = j_{ij}^*$, which implies
\begin{align*}
    \sum_{i<j} j_{ij}^* F_{ij}^\text{cons} \leq \sum_{i<j} j_{ij}^* \, F_{ij}^* + \frac{1}{4} \sum_{i<j} j_{ij}^\text{cons} F_{ij}^\text{cons} 
.\end{align*}
By virtue of Eq. \eqref{eq:app_c:cons_char} applied to $j_{ij} = j_{ij}^*$, the left-hand side is $\sigma^\text{cons}$, whereas the right-hand side can be identified as $\sigma^* + \sigma^\text{cons}/4$. 
Rearranging establishes the upper bound in Eq. \eqref{eq:mainres2:bounds}.}   

\clearpage

\end{document}